\documentclass[twocolumn,showpacs,aps]{revtex4}

\usepackage{graphicx}
\usepackage{dcolumn}
\usepackage{bm}

\begin{document}

\title{Nonextensive aspects of  self-organized scale-free gas-like networks}

\author{Stefan Thurner $^{1,}$}
\email{thurner@univie.ac.at}
\author{Constantino Tsallis$^{2,3,}$}
\email{tsallis@santafe.edu}
\affiliation{
   $^1$Complex Systems Research Group, HNO, 
   Medical University of Vienna, W\"ahringer G\"urtel 18-20, A-1090; Austria \\
   $^{2}$Centro Brasileiro de Pesquisas F\'isicas, Rua Xavier Sigaud 150 22290-180 Rio de Janeiro-RJ; Brasil \\
   $^{3}$Santa Fe Institute, 1399 Hyde Park, Santa Fe, NM 87501; USA 
}



%
%

\begin{abstract}
We explore the possibility to interpret as a 'gas' the dynamical 
self-organized scale-free network recently introduced by Kim et al (2005). 
The role of 'momentum' of individual nodes is played by the degree 
of the node, the 'configuration space' (metric defining distance between nodes) 
being determined by the dynamically 
evolving adjacency matrix. In a constant-size network 
process, 'inelastic' interactions occur between pairs of nodes, 
which are realized by the merger of a pair of two nodes into one. The 
resulting node possesses the union of all links of the previously 
separate nodes. We consider {\it chemostat} conditions, i.e., for each 
merger there will be a newly created node which is then linked to
the existing network randomly. We also introduce an interaction 'potential' (node-merging probability) which decays with distance $d_{ij}$ as $1/d_{ij}^{\,\alpha}\;(\alpha \ge 0)$.  
We numerically exhibit that this system exhibits nonextensive statistics in the 
degree distribution, and calculate how the entropic index $q$ depends on $\alpha$. The particular cases $\alpha=0$ and $\alpha \to \infty$ recover
the two models introduced by Kim et al.
\end{abstract}

\maketitle

\section{{\bf Introduction}}
A common feature in several recent approaches to complex networks with statistical mechanical methods is the 
definition of network Hamiltonians \cite{newman_sm,vicsek,berg04}. 
Such Hamiltonians currently depend on the number of links 
either on a global level, or on the degree of individual nodes. 
This definition allows to borrow powerful concepts from statistical 
physics such as the maximum entropy principle \cite{newman_sm}, 
which may provide the most probable ensembles of networks. 
Further, Hamiltonians allow to define both thermodynamical ensembles 
(microcanonical, canonical, grand canonical) \cite{vicsek,dormen03} and a partition function, which opens the way to compute 
degree correlation functions in a formalism most familiar to physicists \cite{newman_sm,berg04}.

However, these approaches do not yet aim to explain the structure 
of degree distributions, and mainly address random networks. 
A conceptually different approach has been 
taken recently in \cite{tsallis_sm}, where networks are embedded 
in some metric space and the definition of entropy in networks 
is broadened. In this work it was noted that 
the characteristic distribution of the relevant degree of freedom --
the degree of nodes -- appears to coincide to distribution 
functions known for nonextensive systems \cite{tsallis88,gellman}.  
More precisely, in \cite{tsallis_sm} it was found that, for some  
preferential attachment network growth models, the resulting 
degree distributions  are of the {\it $q$-exponential} type (defined later on).
In the usual preferential attachment model, the 
probability of a new node (i.e., being added to a network) to 
attach its link to a pre-existing node $i$ is proportional 
to this node's number of links, or degree $k_i$, i.e., $p_A \propto k_i$. 
This is also true for networks embedded in $R^n$, where the linking 
probabilities are made dependent on the relative distance of node $i$ 
to the new node, i.e., $p_A \propto k_i / d_{ij}^{\,\alpha}$.
Here  $\alpha$ is a free parameter that defines the 
connecting horizon of a new node to the system. 
For large $\alpha$ the node will link with high probability 
to a nearby node, whereas distance becomes irrelevant for 
$\alpha \to 0$.

A problem which has not yet been explored in the literature is 
that of a definition of an interaction between nodes, 
for example in the way one would think of an interaction 
of gas molecules. In classical statistical 
mechanics interactions/collisions between gas particles result in a 
transfer of momentum from one particle to another, under the 
constraint that momentum is conserved. In elastic interactions 
this results in a change of direction and speed of particles 
after a collision, in inelastic ones also in a change of masses 
of the particles.  
In this paper we find that the class of self-organizing networks as introduced in 
\cite{sneppen}
opens the possibility to define an 'interaction' between nodes of a  network. 
In analogy to the momentum transfer in the classical situation, 
in the network case the interactions are defined by a transfer of links. 
This enables one to think of a network as some sort of gas, 
which turns out to be describable by distribution functions characteristic 
of nonextensive statistical mechanics.

\section{{\bf Model}}

Let us consider the following gas-like system. Particles have links among them. 
The total number of links of a given node is a characteristic
quantity of the node, such as the momentum of an ordinary particle.
The particles of this 'gas' have no momentum but only their degree.
Neither do the particles have an absolute position in space. They possess only a 
relative distance $d_{ij}$ to each other which is given by the {\it shortest number of 
links} between them (sometimes called {\it chemical distance}). 
Particles can interact 'non-elastically'. Upon an 
interaction one particle ceases to exist and transfers all its links to the 
other. If the interacting particles $i$ and $j$ have both had links to a 
third particle $k$ before the interaction, the remaining particle $i$ 
will keep its link to $k$, while the links of the disappeared particle $j$ to $k$
will be removed, meaning that links are only counted once (and are not weighted).
Consider these interactions taking place in a {\it chemostat}, 
such that the number of particles in a closed system is constant.
This means that, for every merging
 interaction, a new particle will be introduced to 
the system. 
The interaction is characterized by the probability that two particles 
meet and transfer links. Given a 'metric' (relative network distance) 
this  probability can be made distance-dependent  as in 
\cite{tsallis_sm}. To do this we introduce a power like 
potential.

In what follows we numerically study the distribution of the 
characteristic degree of freedom (the degree of nodes) as a function of
the range of the interaction. As in \cite{tsallis_sm}, we find that 
the distribution of this (nonextensive) 'gas' is  described by a $q$-exponential.
\begin{figure}[htb]
\begin{center}
\begin{tabular}{c} 
\includegraphics[height=63mm]{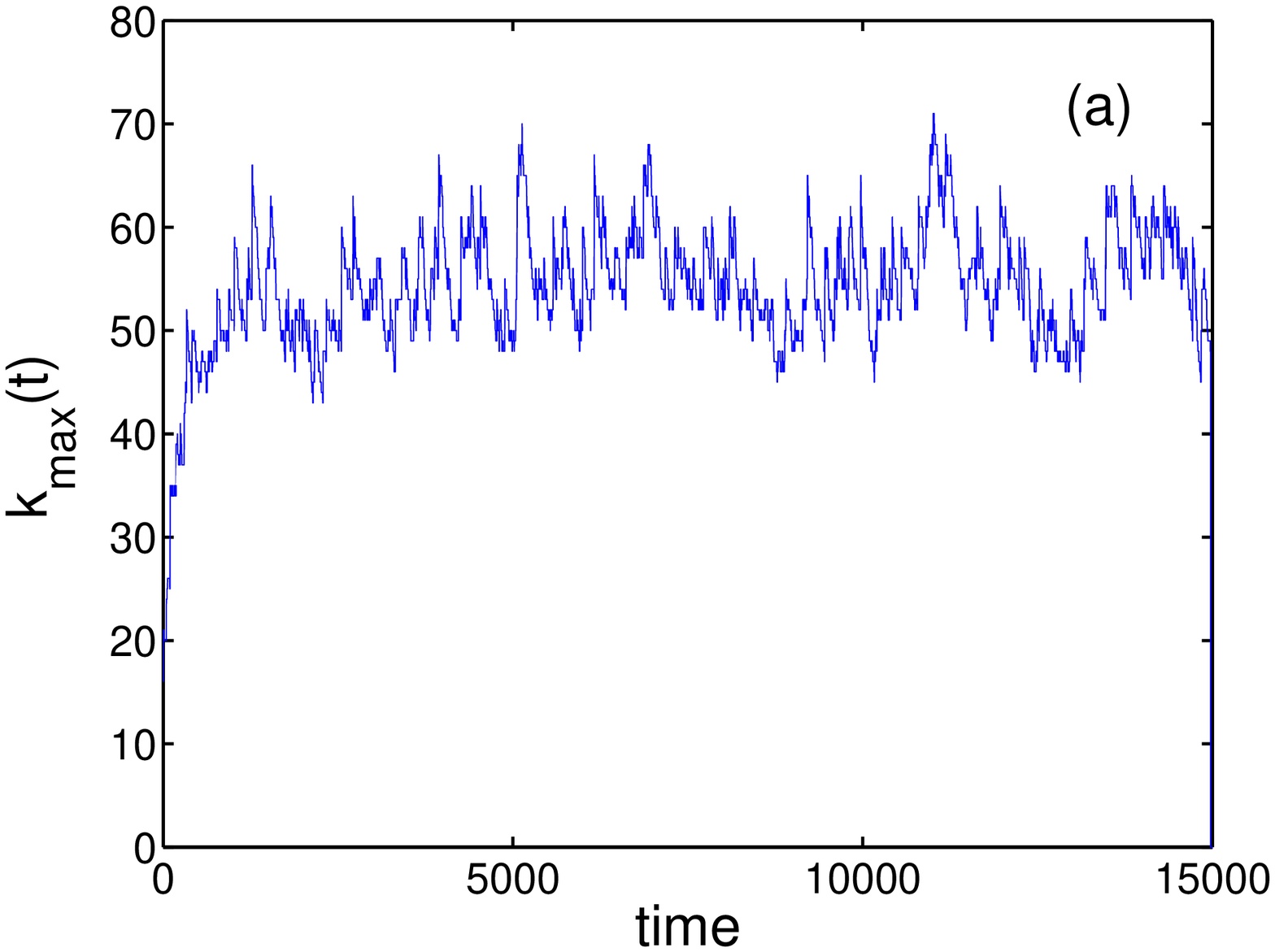} \\
\includegraphics[height=63mm]{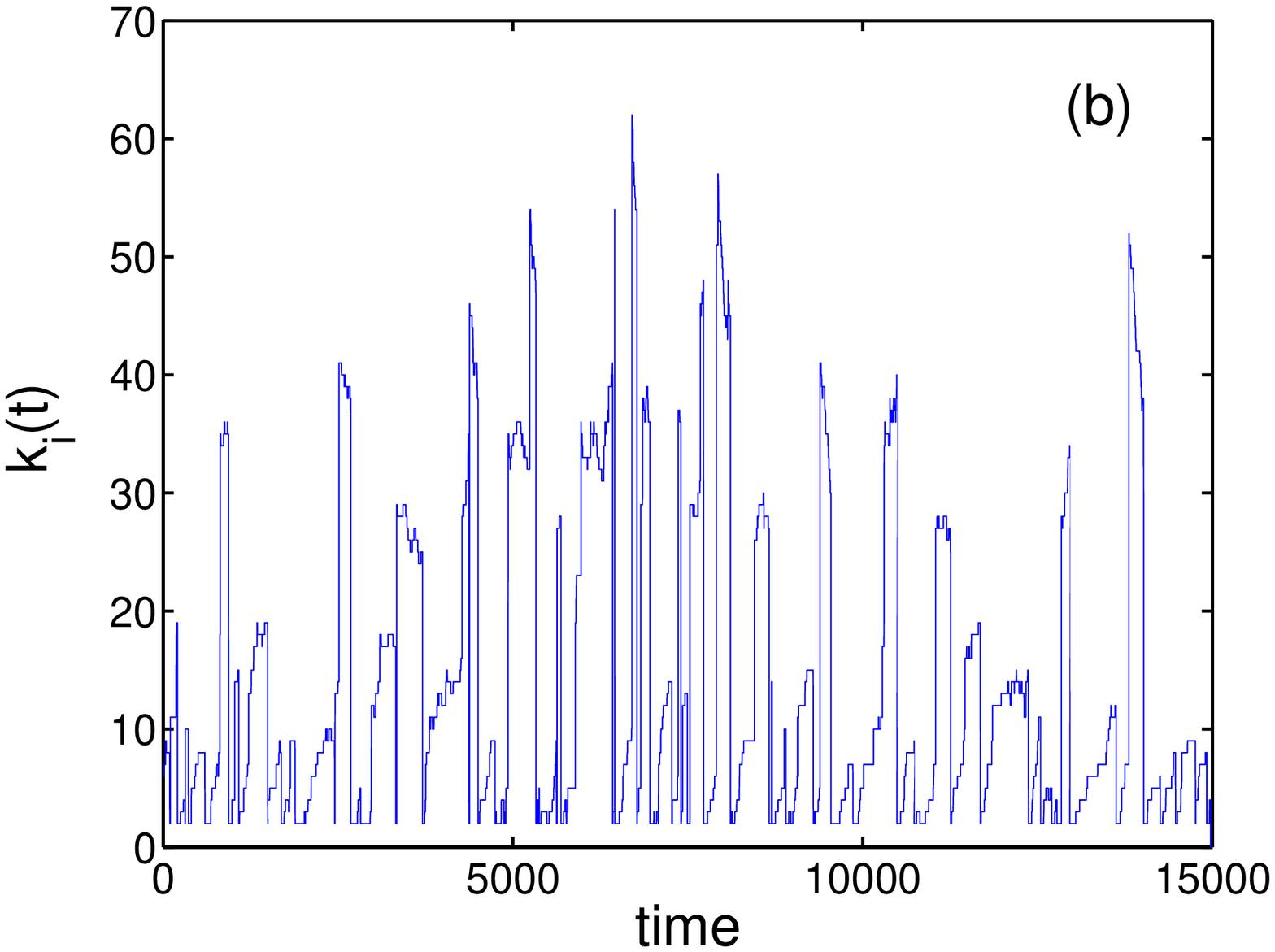} \\
\end{tabular}
\end{center}
\caption{Simulation of Eq. (\ref{update}) of a $N=2^7$ network with $r=2$ and $\alpha=0$ (random scheme).
Time evolution of the degree of (a) the most connected node during the simulation; (b) an individual, randomly chosen node.
}
\label{dyn} 
\end{figure}

\subsection{Network dynamics}
\begin{figure}[tb]
\begin{center}
\begin{tabular}{c} 
\includegraphics[height=65mm]{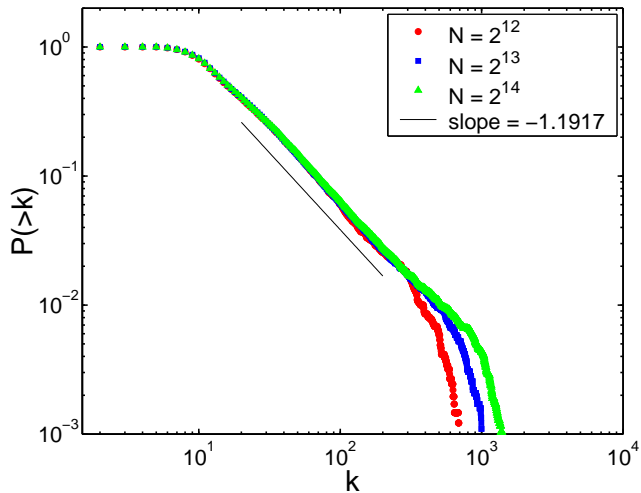} \\
\end{tabular}
\end{center}
\caption{Cumulative degree distributions for various system sizes $N$ for
the neighbor scheme, as reported in \cite{sneppen} with $\langle r \rangle=8$.
Degrees are from individual states of networks without averaging over 
identical realizations. On the right side the exponential finite size 
effect is seen. 
}
\label{deg} 
\end{figure}

There has been a recent suggestion to model scale-free networks of 
constant size, the so-called self-organized networks \cite{sneppen}.
The idea is that at any given timestep 
one single pair of nodes, say $i$ and $j$, merge together 
to become one single node. This node keeps the name of one 
of the original nodes, say for example $i$. This node now gains 
all the links of the other node $j$, resulting in a 
change of degree for node $i$  according to 
\begin{eqnarray}
k_i &\to& k_i + k_j -N_{\rm common},   {\rm if \,\,} (i,j){\rm  \,\, not \,\, first \,\, neighbors} \nonumber \\
k_i &\to& k_i + k_j -N_{\rm common}-2, {\rm if \,\,} (i,j){\rm  \,\, first \,\, neighbors} \nonumber \\
\label{update}
\end{eqnarray}
where $N_{\rm common}$ is the number of nodes, which shared 
links to both of $i$ and $j$ before the merger. In the case that 
$i$ and $j$ were first neighbors before the merger, i.e., they had been previously linked,  
the removal of this link will be taken care of by the term $- 2$ in Eq. (\ref{update}).
Next, to keep the system at constant size, a new node is created, 
and is linked to $r$ randomly 
chosen nodes from the existing network. Let us note here that 
the smallest degree found in a network can only be larger or equal 
to $r$. This will have consequences for the normalization of 
distribution functions as will be discussed
below. In the following (except for Fig. \ref{deg}) we will restrict 
ourselves to $r=2$, for simplicity. 
Thus the smallest degree will always be 2.
(This is not at all an important restriction; as an alternative the 
actual number of links can also be a random number picked, e.g. from a uniform 
distribution with an average of $\langle r \rangle$, as in \cite{sneppen}).
After that we address the next timestep.
Nodes in the network start with a small number of links, and gain 
links through merger-interactions. Links to the $N_{\rm common}$
common neighbors of two merging nodes are lost as mentioned above, 
which reduces the number of links.    
Gains and losses eventually lead to an effective balance over time as   
shown for instance in Fig. \ref{dyn} a. The number of links of the best 
connected node in the system is followed over time. After 
about 1000 timesteps a stationary state is reached. The situation for 
an individual link is shown in Fig. \ref{dyn} b. A node starts 
with $r=2$ links when it is introduced to the system. It gains 
links through mergers over time. When the node is taken up in a 
merger it disappears from the system and, as said before, a new one starts with $r=2$ links
again. 
Networks with these rewiring scheme lead to scale-free degree 
distributions \cite{sneppen}, i.e. the power exponent of the cumulative 
degree distribution tail behaves as, $P( \ge k) ~ \sim k^{-\gamma}$.
 In \cite{sneppen} two schemes were 
discussed: The case where only nodes being {\it first-neighbors}  
can merge, and the case where {\it any} two nodes -- directly connected or not -- 
can merge with the same probability for each possible pair $(i,j)$.
The {\it neighbor scheme} leads to an exponent $\gamma \sim 1.3$, the {\it random scheme} 
to $\gamma \sim 0.5$. The cumulative degree distribution,
for the neighbor scheme is 
shown in Fig. \ref{deg}.

These distributions can be fit by q-exponentials, 
\begin{equation}
P( \ge k)= e_{q_c}^{-(k-2)/\kappa} \;\;\;\;(k =2,3,4,...)\quad ,
\label{ans}
\end{equation}
where the $q$-exponential function is defined, for $1+(1-q_c)x \ge 0$, as 
\begin{equation}
e_{q_c}^x \equiv [1+(1-q_c)x]^{1/(1-q_c)} 
\end{equation}
with $\kappa>0$ some characteristic number of links, and 
$\gamma \equiv 1/(q_c-1)$ being the  tail exponent of the (asymptotic) power-law 
distribution. 
Whenever we talk about q-values corresponding to a cumulative distribution, 
we  use the notation $q_c$, where $c$ stands for {\it cumulative}.  
We have normalized with the value corresponding to the smallest possible 
degree (which in our case equals 2) in order to have $P( \ge 2)=1$.

\begin{figure}[tb]
\begin{center}
\begin{tabular}{c} 
\includegraphics[height=65mm]{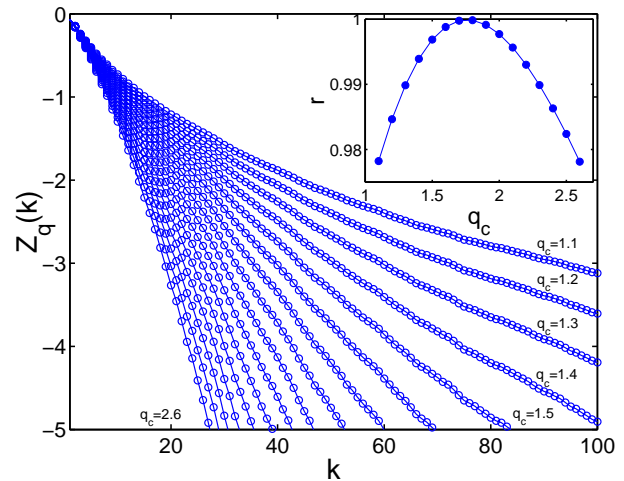}
\end{tabular}
\end{center}
\caption{$q$-logarithm of the cumulative distribution function 
from the previous figure as a function of the degree $k$. Clearly, 
there is a $q_c$ which allows for an optimal linear fit.  
{\it Inset:} Linear correlation coefficient $r$ of 
$Z_q(k)$. The value of 
$q_c$ is obtained where $Z_q$ is optimally linear, i.e., where the 
correlation coefficient is maximal. In this example we obtain $q_c=1.84$, 
which corresponds to the slope $\gamma=1.19$ in the previous plot. 
The continuous curves are guides to the eye. }
\label{Z} 
\end{figure}

A convenient procedure to perform a two-parameter fit of this kind is 
to take the {\it $q$-logarithm} of the distribution $P$, defined by
$Z_{q} \equiv \ln_{q} P(\ge k) \equiv  \frac{[P(\ge k)]^{1-q_c}-1}{1-q_c}$. 
This is done for a 
series of different values of $q_c$. The function $Z_{q}$ which is 
best fit with a straight line determines the value of $q_c$, the  
slope being  $-\kappa$. The situation for the $N=2^{14}$ data of Fig. \ref{deg}
is shown in Fig. \ref{Z} for $q_c$ running between 1 and 2.6. 

We numerically verify with good precision that the {\it Ansatz} in Eq. (\ref{ans}) for the cumulative degree distribution is a satisfactory one (it could even be the {\it exact} one for the present problem). This reveals a connection \cite{tsallis_sm,doye} of scale-free network dynamics to  nonextensive statistical mechanics \cite{tsallis88,gellman}. Let us be more specific.
Consider the entropy 

\begin{eqnarray}
S_q &\equiv& \frac{1-\int_2^\infty dk \, [p(k)]^q}{q-1}  \nonumber \\ 
\Bigl[S_1&=&S_{BG} \equiv -\int_2^\infty dk \, p(k) \ln p(k)\Bigr] \, ,
\end{eqnarray}
where we assume $k$ as a continuous variable for simplicity, and $BG$ stands for {\it Boltzmann-Gibbs}. If we extremize $S_q$ with 
the constraints \cite{TsallisMendesPlastino}
\begin{equation}
\int_2^\infty dk \, p(k)=1 
\label{constr1}
\end{equation}
and
\begin{equation}
 \frac{\int_2^\infty dk \, k \, [p(k)]^q}{ \int_2^\infty dk \,  [p(k)]^q  } = K\,,
 \label{constr2}
\end{equation}
we obtain
\begin{equation}
p(k)= \frac{e_q^{-\beta (k-2)}}{\int_2^\infty dk^\prime  \, e_q^{-\beta (k^\prime-2)}}=\beta(2-q) e_q^{-\beta (k-2)} \;\;\;\;(k \ge 2) \,,
\end{equation}
where $\beta$ is determined through Eq. (\ref{constr2}). Both 
positivity of $p(k)$ and normalization constraint (\ref{constr1}) impose $q<2$.
The corresponding cumulative distribution $P(>k)$ is then given by
\begin{equation}
P(>k) \equiv 1 - \int_2^k dk^\prime \, p(k^\prime) = [1-(1-q)\beta (k-2)]^{\frac{2-q}{1-q}} \,.
\end{equation}
This expression can be rewritten precisely as the Ansatz (\ref{ans}) with 
\begin{equation}
q_c \equiv \frac{1}{2-q} \,; \;\kappa \equiv \frac{1}{(2-q)\beta} \,.
\end{equation}

\begin{figure}[tb]
\begin{center}
\includegraphics[height=65mm]{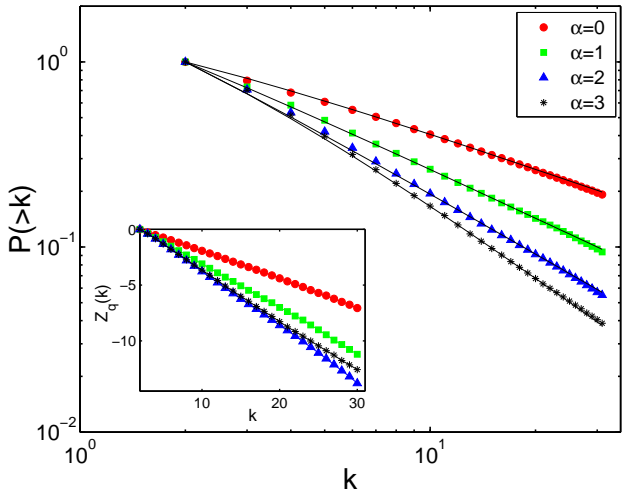}
\end{center}
\caption{Cumulative degree distribution for a $N=2^9$ network with $r=2$, for typical  
values of $\alpha$. For $k>30$ finite size effects (such as those illustrated in 
Fig. 2) emerge. Solid lines are the  $q$-exponentials from 
the fitted parameters $q_c$ and $\kappa$, shown in Fig. \ref{qalpha} a. {\it Inset:} the same data in 
{\it $q$-logarithm vs. linear representation}, where we have used, for each value 
of $\alpha$ the corresponding value of $q_c$. For $\alpha$ ranging within the interval 
$[0,7]$ we verify that the corresponding linear correlation 
coefficient ranges within the interval $[ 0.999901, 0.999976]$. The slope equals 
$-\kappa$ for each curve (see also inset of Fig. \ref{qalpha} a). 
}
\label{deg_dep} 
\end{figure}

\subsection{Network distance and distance-dependent re-linking potential}

Unlike the two schemes in the original presentation of self-organizing networks, 
the neighbor and the random scheme, we would like to define a distance-dependent 
merging probability. This needs a definition of {\it distance} on the graph. 
For simplicity we define the distance $d_{ij}$, between two nodes $i$ and $j$ 
on an undirected graph as the shortest distance, given all links are of unit length.  
This distance is obtained, for instance, from the Dijkstra algorithm \cite{dijk}.
We randomly choose a node (denoted by $i$) and then choose the second node (denoted by $j$) of the merger with  probability 
\begin{equation}
p_{ij}= {\cal N} d_{ij}^{-\alpha}  \;\;\;\;(\alpha \ge 0) \quad , 
\end{equation}
where ${\cal N} = 1/\sum_j d_{ij}^{-\alpha}$ is a normalization that makes $p_{ij}$ a probability, 
and $d_{ij}$ is the shortest distance (path) on the network 
connecting nodes $i$ and $j$; $\alpha$ is a real number.
Obviously, tuning $\alpha$ from $0$ toward large values 
switches the model from the random  to the neighbor scheme in \cite{sneppen}.  

\begin{figure}[t]
\begin{center}
\includegraphics[height=65mm]{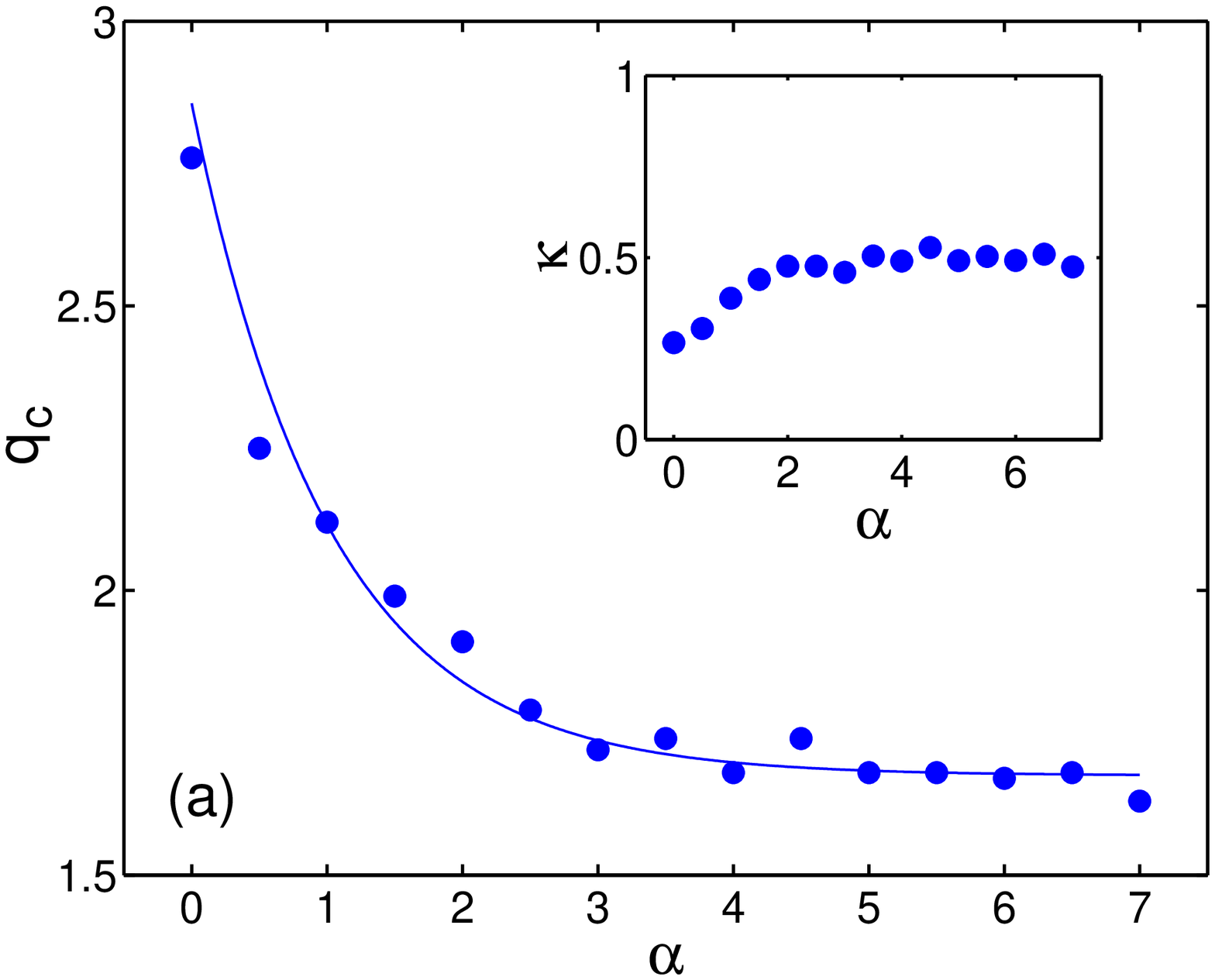} \\
\includegraphics[height=65mm]{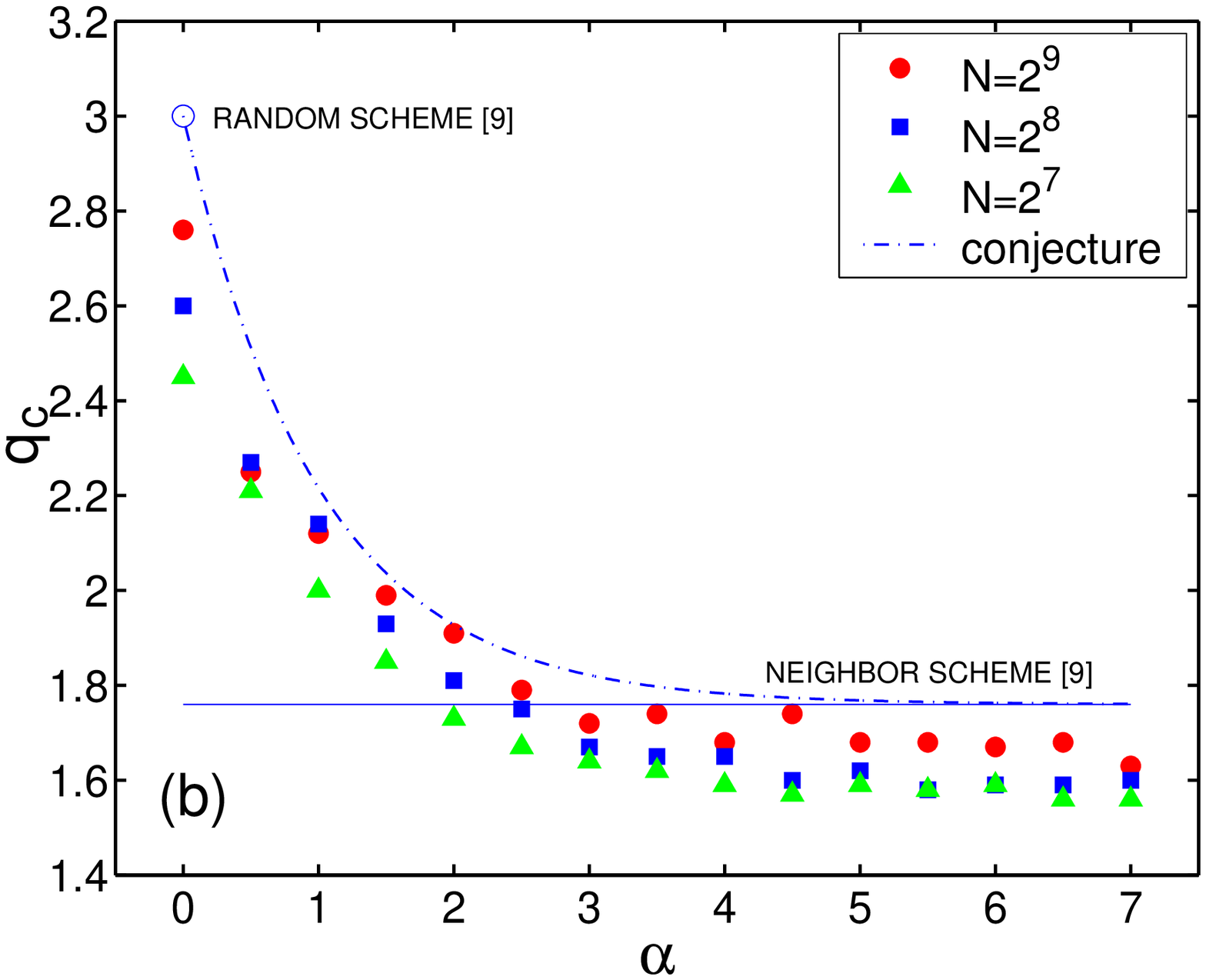}
\end{center}
\caption{(a) Dependence of the extensivity parameter $q_c$ on the 
potential parameter $\alpha$ for a $N=2^9$, $r=2$, network. The solid line is an exponential 
fit to the data points: $q_c=1.675+ 1.1809 \; e^{-0.985 \, \alpha}$. 
We might thus conjecture for the exact answer, the form $q_c(\alpha)=q_c(\infty) + [q_c(0)-q_c(\infty)] \, e^{-\lambda \alpha}$ with $\lambda \simeq 1$. 
{\it Inset:} $\alpha$- dependence of $\kappa$. 
(b) Network size dependence of the $q_c(\alpha)$ function, for 
nodes $N=2^7,2^8,2^9$. The solid line corresponds to the asymptotic 
value for large $\alpha$ reported in \cite{sneppen} (neighbor scheme), i.e., $q_c(\infty)=1.76$. 
The value for the $\alpha=0$ model as reported in \cite{sneppen} corresponds 
to $q_c(0)=3$ (empty circle). The conjectural form is indicated  with $\lambda =1$ (dashed line).
}
\label{qalpha} 
\end{figure}

\section{{\bf Results}}
Realizing this distance-dependent potential in a numerical 
simulation we find the degree distributions given in Fig. \ref{deg_dep}.
All following data was obtained from averages over 1000 identical 
realizations of degree distributions 
of networks with a number of nodes, $N=2^9$.
Networks have been recorded after 5 network updates. A network update 
is performed when $N$ mergers have taken place. 
This corresponds to the $5\times 2^9$ timesteps shown in Fig. \ref{dyn}.

From these degree distributions we obtain the index $q_c$ and the characteristic degree $\kappa$. 
Their dependence on $\alpha$ is given in Fig. \ref{qalpha}. 
The $\alpha$-dependence of $q_c$ shows the expected tendency. 
Our result in the limit $\alpha \to 0$, $q_c(0)=2.8$
corresponds to an exponent $\gamma = 0.55$, 
which is about 10 percent lower than the reported value in \cite{sneppen}. 
The reason for this small discrepancy seems to be a finite-size effect.

To demonstrate that this might indeed be so,  in Fig. \ref{qalpha} b we plot the 
$q_c(\alpha)$-dependence for $N=2^7$, $N=2^8$ and $2^9$ networks 
for comparison. As network size increases the value of $q_c$ approaches 
the expected value of $3$ \cite{sneppen} in the small $\alpha$ limit. 
For the $\alpha \to \infty$ limit, the expected 
value is recovered for the $N=2^9$ network, for smaller 
networks, there is still a visible size effect. 
This size effect is related to the problem that the finite 
size cutoff plays a relatively large role, and interferes 
considerably in the fits in small networks. Simulations on larger networks are certainly desirable, but inaccessible to our present computational power.

\section{{\bf Discussion}}

To summarize, we have explored the possibility to making some connection between a nonextensive 
gas and a self-organized scale-free network. We have shown that the 
characteristic degree distributions are well described by q-exponentials 
whose parameters vary with the interaction range, i.e. $\alpha$.
The limiting cases  $\alpha \to 0$ and $\alpha \to \infty$ reproduce 
the situations given in \cite{sneppen}, namely the neighbor merging  
and the random merging.

In the present work we have used the network's intrinsic metric space, i.e., its 
adjacency matrix, to measure distances ($d_{ij}$) between nodes to be merged. This is in
variance with what is done in  \cite{tsallis_sm}, where the network  is embedded  in 
a 'geographical' metric space (with distances $r_{ij}$). Both models can of course be unified by introducing both merging (with probability $\propto 1/d_{ij}^{\, \alpha}$) and distance-dependent linking (with probability $\propto 1/r_{ij}^{\alpha_A}$, where $A$ stands for {\it attachment}). The degree distribution of such a unified model could still be of the $q$-exponential form with $q_c(\alpha,\alpha_A)$. Of course,  $q_c(\alpha,0)=q_c(\alpha)$ as given in the present paper. It would not be surprising if $q_c(\alpha,\alpha_A)$ was a monotonically decreasing function of both variables $\alpha$ and $\alpha_A$, with the maximal value being $q_c(0,0)$, and with say  $q_c(\alpha,\infty)=1,\,\forall \alpha>0$. In such a case, the interval spanned by $q_c$ would clearly be wider than that of the present model. 

S.T. would like to thank the SFI and in particular J.D. Farmer for their great hospitality and support 
during Sept-Oct of 2004, when this work was initiated. 
Support from SI International and AFRL/USA is acknowledged as well.

\end{document}